\documentclass[prd,twocolumn,showpacs,preprintnumbers,nofootinbib,amsmath,amssymb,nobalancelastpage]{revtex4}

\usepackage[pdftex]{graphicx}
\usepackage{latexsym,amsmath,amssymb,lmodern,float,url}
\usepackage{natbib}
\usepackage[pdftex,bookmarks,linktocpage,pdfpagelabels,plainpages=false,hyperfigures,linkcolor=blue,citecolor=blue,urlcolor=blue]{hyperref} 
\usepackage{xspace}
\usepackage{courier}
\hypersetup{colorlinks=true}
\usepackage{bm}
\usepackage{dcolumn}
\usepackage{amsmath}
\usepackage{amssymb}
\usepackage{color}
\usepackage{xcolor}
\usepackage{soul}
\usepackage{url}
\usepackage{float}

\usepackage{aas_macros}
\usepackage[normalem]{ulem}
\usepackage[utf8]{inputenc}
\usepackage{float}

\def\Eq#1{Eq.~(\ref{#1})}
\def\Fig#1{Fig.~(\ref{#1})}

\def\Tab#1{Tab.~\ref{#1}}

\def\H0{$H_0$}
\def\si8{$\sigma_8$}

\begin{document}

\title{Deep learning the astrometric signature of dark matter substructure} 

\author{Kyriakos Vattis}
\email{kyriakos\_vattis@brown.edu}
\affiliation{ Department of Physics, Brown University, Providence, RI 02912-1843, USA}
\affiliation{ Brown Theoretical Physics Center, Brown University, Providence, RI 02912-1843, USA}

\author{Michael W. Toomey}
\email{michael\_toomey@brown.edu}
\affiliation{ Department of Physics, Brown University, Providence, RI 02912-1843, USA}
\affiliation{ Brown Theoretical Physics Center, Brown University, Providence, RI 02912-1843, USA}

\author{Savvas M. Koushiappas}
\email{koushiappas@brown.edu}
\affiliation{ Department of Physics, Brown University, Providence, RI 02912-1843, USA}
\affiliation{ Brown Theoretical Physics Center, Brown University, Providence, RI 02912-1843, USA}

\date{\today}

\begin{abstract}
We study the application of machine learning techniques for the detection of the astrometric signature of dark matter substructure. In this proof of principle a population of dark matter subhalos in the Milky Way will act as lenses for sources of extragalactic origin such as quasars. We train {\it ResNet-18}, a state-of-the-art convolutional neural network to classify angular velocity maps of a population of quasars into lensed and no lensed classes. We show that an SKA -like survey with extended operational baseline can be used to probe the substructure content of the Milky Way, and demonstrate how axiomatic attribution can be used to localize substructures in lensing maps.

\end{abstract}

\maketitle
\section{Introduction}
The Standard Cosmological Model, $\Lambda$CDM consisting of 70$\%$ dark energy in the form of a cosmological constant, 25$\%$ of Cold Dark Matter and 5$\%$ of baryonic matter has been quite successfully tested in recent years from galactic scales to cosmological scales \cite{2018arXiv180706209P, 2019MNRAS.483.4866A, 2018MNRAS.477.1604G,2000MNRAS.311..441C,2008ApJ...679.1173R}.  However one of its main components, dark matter, has only been observed through its gravitational effects despite all the efforts through  direct and indirect detection \cite{Drukier:1986tm,Goodman:1984dc,Akerib:2016vxi,Cui:2017nnn,Aprile:2018dbl,2020arXiv200304545F,2015JCAP...09..008F,2015PhRvD..91h3535G,2018ApJ...853..154A,2017PhRvD..95h2001A,2020Galax...8...25R,2016JCAP...02..039M,2017arXiv170508103I,2015arXiv150304858T}  and colliders like the Large Hadron Collider \cite{Aaboud:2019yqu,2017JHEP...10..073S} to observe Weakly Interacting Massive Particles (WIMP) or other popular candidates like axions \cite{2018PhRvL.120o1301D, 2015ARNPS..65..485G,2020arXiv200610735K,2020arXiv200612488B}. While the efforts continue, the null results highlight the necessity of finding alternative gravitational signatures that would shed light on the  elusive nature of dark matter.

A very promising way to probe various dark matter models  gravitationally is to study their effects on the distribution of dark matter on subgalactic scales, an idea that has been studied thoroughly in the literature. Several methods have been proposed including the utilization of tidal streams \cite{2014ApJ...788..181N,2016ApJ...820...45C,2016PhRvL.116l1301B,2016MNRAS.463..102E,2016arXiv160805624S, 2019ApJ...880...38B}, astrometric observations \cite{2015MNRAS.446.1000F,2018JCAP...07..041V,2020arXiv200302264M, 2020PhRvL.125k1101M}, pulsar timing array observations \cite{2019PhRvD.100b3003D}, stellar wakes \cite{2018PhRvL.120u1101B} and images of strongly-lensed background galaxies or quasars \cite{2010ApJ...724..400C,2011ApJ...741..117C, 2018PhRvD..97b3001D,2018PhRvD..98j3517D,sub1,sub2,sub3,alma,veg,koop,veko,2010MNRAS.408.1969V,2012Natur.481..341V,2014MNRAS.442.2017V,Ritondale:2018cvp} looking for dark subhalos. For most of the aforementioned methods the expected signal is small and difficult to observe.  

With the increase of computational power and the improvement of machine learning algorithms there have been multiple attempts to involve neural network--based approaches for such analyses. 
In the last few years machine learning has been used in an array of problems in cosmology \cite{Ntampaka:2019udw}  as well as in other problems in the physical sciences \cite{Carleo:2019ptp}.  
Examples include applications in large-scale structure \cite{ Pan:2019vky}, the Cosmic Microwave Background \cite{Mishra:2019mbm,Farsian:2020adf, Caldeira:2018ojb}, the cosmological 21~cm signal  \cite{Hassan:2019cal} and lensing studies \cite{ 2010ApJ...720..639G, Nurbaeva:2014fsa,Schmelzle:2017vwd,Fluri:2019qtp, Hezaveh:2017sht,PerreaultLevasseur:2017ltk,Morningstar:2018ase,Morningstar:2019szx,canameras2020holismokes}. 
In the search for dark matter, supervised machine learning algorithms have been trained to identify substructure properties with simulated galaxy-galaxy strong lensing images. 
More recently, convolutional neural networks (hereafter, CNN) have been used for classification of different subhalo mass cut-offs \cite{Varma:2020kbq}, classifying DM halos with and without substructure \cite{Alexander:2019puy, DiazRivero:2019hxf}, to distinguish dark matter models with disparate substructure morphology \cite{Alexander:2019puy}, and also inference of population level properties for DM substructure \cite{Brehmer:2019jyt}.

In this work we investigate the possibility of applying machine learning techniques and in particular a CNN, to detect the presence of lensing effects from dark matter substructure on a population of quasars in astrometric data from future surveys based on simulated data as a proof of principle. We also demonstrate how axiomatic attribution can be used to localize substructures in lensing maps. In Sec.~\ref{sec:formalism} we outline the formalism followed in order to simulate the lensing signal followed by a description of creating simulated data maps. We discuss the implementation of the CNN in Sec.~\ref{sec:DL}, our results in Sec.~\ref{sec:results} and our conclusions are shown in Sec.~\ref{sec:conclusion}.

\section{Astrometric lensing} \label{sec:formalism}
\subsection{Formalism}
We begin by considering the lensing effect of a dark matter halo on the astrometric parameters of a background source (position, velocity and acceleration), similar in spirit to \cite{2018JCAP...07..041V,2020arXiv200302264M}. We utilize a modified version of the code provided in \cite{2020arXiv200302264M} to extract the astrometric signal in a suite of models that are used to create the simulated data set used in deep learning model training.  

\begin{figure}
\centering
\includegraphics[width=0.99\columnwidth]{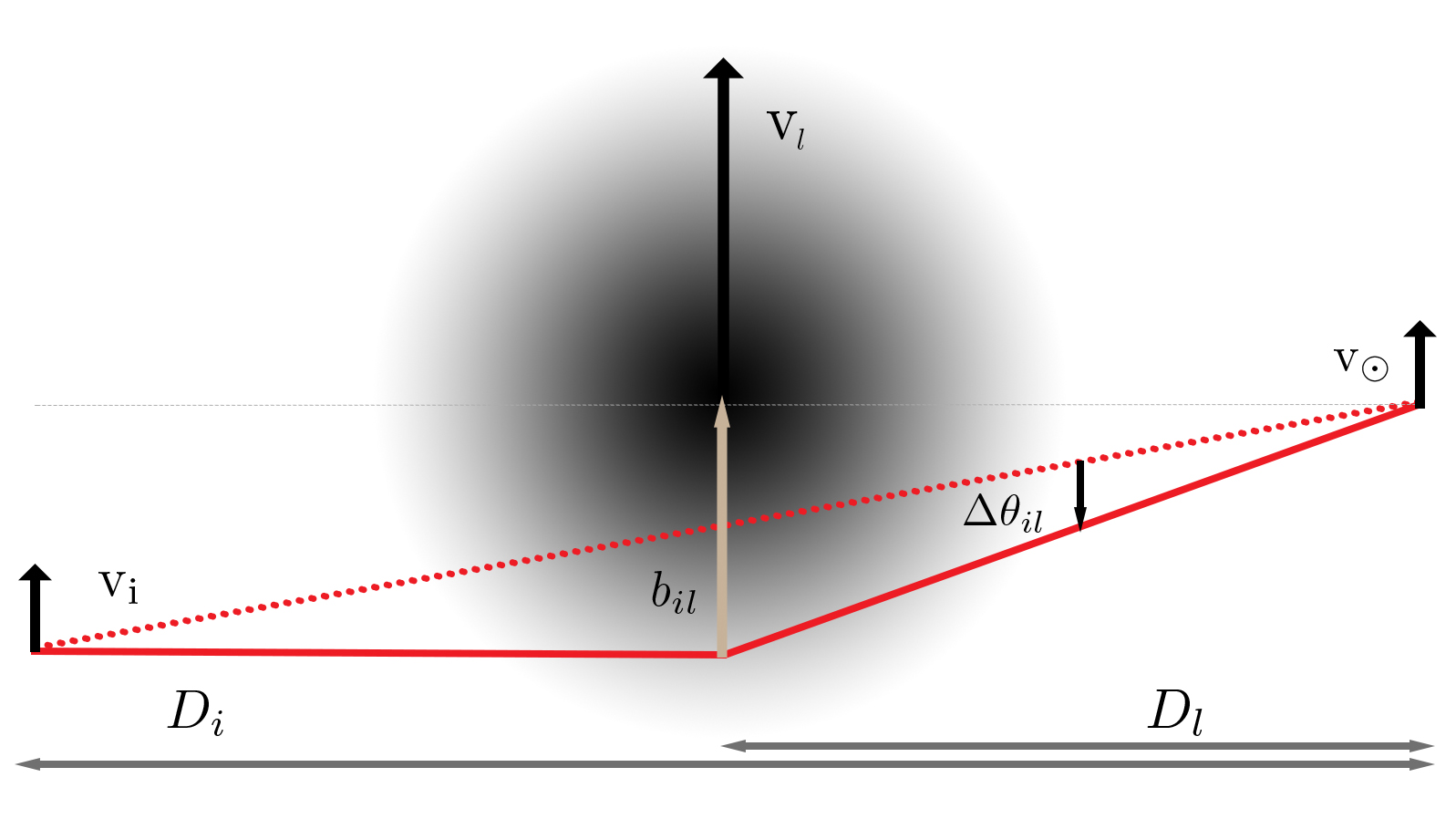}  
\caption{Astrometric lensing of a source $i$ by a lens $l$ moving with velocities $\mathbf{v}_i$ and $\mathbf{v}_l$ respectively as observed by an observer moving with velocity $\mathbf{v}_\odot$. The impact parameter vector $\mathbf{b}_{il}$ is shown in gold and and $\mathbf{ \Delta \theta} _{il}$ is the induced angular deflection. All vectors are two dimensional and the distances $D_l$ and $D_i$ are considered constant.}
\label{fig:fig1}
\end{figure}

The system we consider is shown in \Fig{fig:fig1}; a source i moving with velocity $\mathbf{v}_i$ is lensed by a lens $l$ with velocity $\mathbf{v}_l$ as seen by an observer with velocity $\mathbf{v}_\odot$. The distance from observer to a lens is $D_l$ and from observer to a source is $D_i$. Both are assumed to be constant and thus all vectors involved in the calculation are two dimensional laying on the celestial sphere.

\begin{figure*}
\centering
\includegraphics[width=1.95\columnwidth]{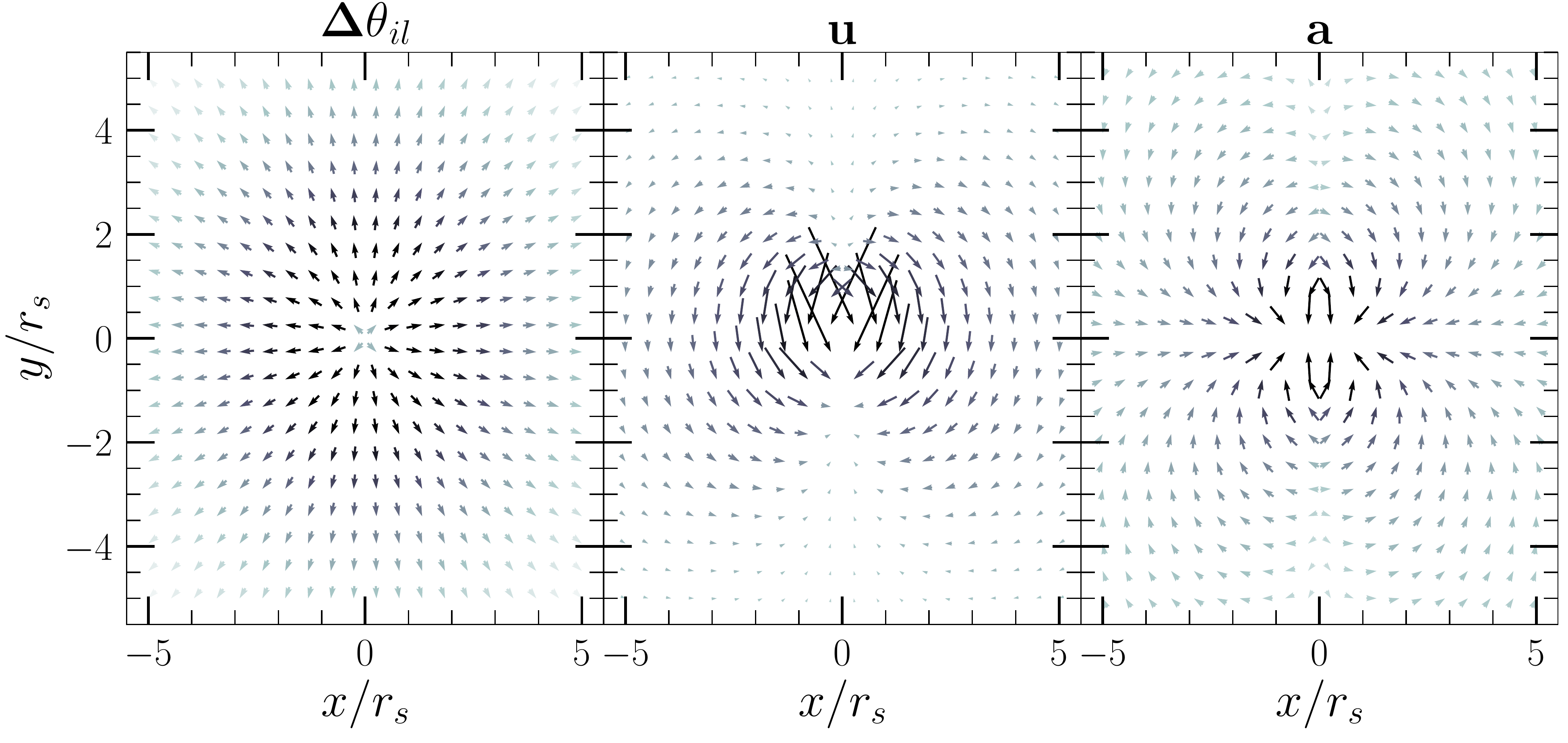}  
\caption{The angular deflection $\mathbf{ \Delta \theta} _{il}$ and its first two time derivatives in arbitrary units induced by an NFW halo centered on the origin and moving in the $+ \hat{\bf{y}}$ direction.  Darker colors correspond to the magnitude of the vector.}
\label{fig:fig2}
\end{figure*}

For such a system the deflection angle $\mathbf{ \Delta \theta} _{il}$ between the real position and the position of the image of the source is given by

\begin{equation} 
\mathbf{ \Delta \theta} _{il} = - \left ( 1- \frac{D_l}{D_i} \right ) \frac{4G_N M(b_{il})}{c^2b_{il}}\hat{\mathbf{ b}}_{il}
\label{eq:deflang}
\end{equation} 
where $G_N$ is Newton's Gravitational constant, $c$ is the speed of light, $\mathbf{ b}_{il}$ the impact parameter vector pointing towards the center of the lens and $M(b_{il})$  the enclosed lens mass which given a spherically symmetric density profile $\rho(r)$  for the lens can be calculated as 

\begin{equation} 
M(b_{il}) = 2 \pi \int ^{+\infty}_{-\infty}dx \int ^{b_{il}}_{0} db^{\prime} b^{\prime} \rho\left(\sqrt{x^2 + b^{\prime2} }\right).
\label{eq:Mencl}
\end{equation}

Because all the components of the system are non stationary, the impact parameter vector $\mathbf{ b}_{il}$ is itself time dependent and as a consequence the deflection vector $\mathbf{ \Delta \theta} _{il}$ has a time dependence as well. To calculate the lensing induced velocity we take the time derivative of \Eq{eq:deflang} resulting to

\begin{multline}
\mathbf{u}=\dot{\mathbf{\Delta \theta}} _{il} =- \left ( 1- \frac{D_l}{D_i} \right ) \frac{4G_N}{c^2} \left \{ \frac{M^{\prime}(b_{il}) \dot{|b_{il}|}}{b_{il}}\hat{\mathbf{ b}}_{il} \right.\\
\left. +  \frac{M(b_{il})}{b_{il}^2} \left[\mathbf{v}_{il} - 2 \dot{|b_{il}|} \hat{\mathbf{ b}}_{il} \right] \right\} 
\label{eq:vel}
\end{multline}
with $M^{\prime}$ the derivative of the enclosed mass with respect to the impact factor. The quantity  $\mathbf{v}_{il}$ is the effective velocity of the lens defined as
\begin{equation} 
\mathbf{v}_{il} = \mathbf{v}_{l} - \left ( 1- \frac{D_l}{D_i} \right )\mathbf{v}_{\odot} - \frac{D_l}{D_i} \mathbf{v}_{i}
\label{eq:effvel}
\end{equation}
where $\mathbf{v}_{l}$ is the velocity vector of the lens, $\mathbf{v}_{i}$ the velocity vector of the source and $\mathbf{v}_{\odot}$ is the velocity vector of the observer. The time derivative of the magnitude of the impact factor is given by
\begin{equation} 
 \dot{|b_{il}|} = \hat{\mathbf{ b}}_{il} \cdot \mathbf{v}_{il}.
\label{eq:bmagder}
\end{equation}

\begin{figure*}
\centering
\includegraphics[width=2\columnwidth]{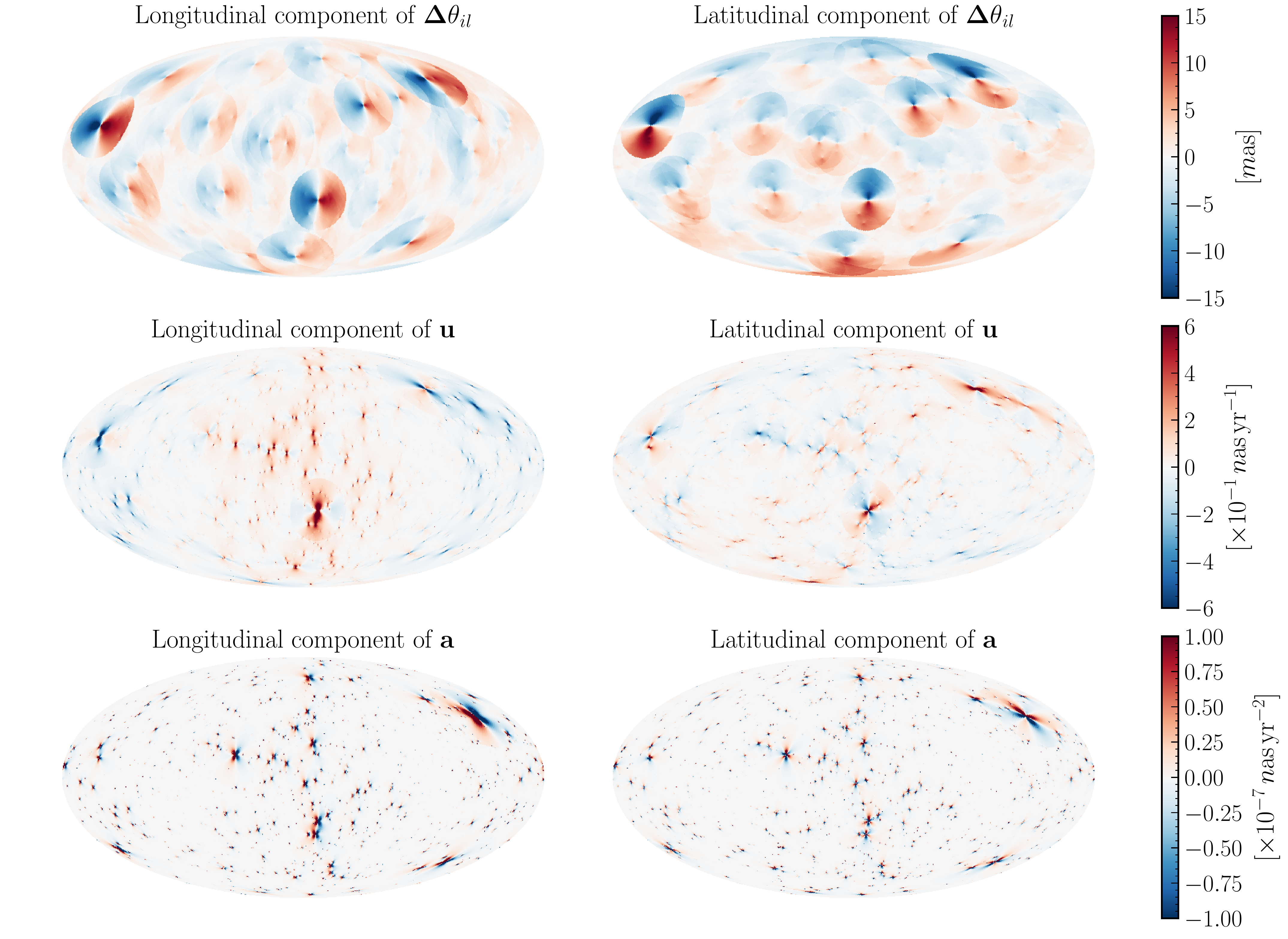}  
\caption{The induced astrometric lensing effect on a population of quasar sources by a population of NFW subhalos distributed in the Milky Way, under the assumption of pure signal and no instrumental noise. We truncate the effect within 20 degrees of the centre of each subhalo. The calculation was done on a HEALPIX map with ${\tt{nside}} = 128$ and we simulate $\sim1200$ subhalos with masses between $10^7$ and $10^{10} M_{\odot}$. The first row shows the angular displacement $\mathbf{\Delta \theta}_{il}$, the second row the induced velocity $\mathbf{u}$ and finally the third row the induced acceleration $\mathbf{a}$. }
\label{fig:fig3}
\end{figure*}

Similarly we take the time derivative of \Eq{eq:vel} to calculate the lensing induced acceleration as
\begin{multline}
\mathbf{a} = \ddot{\mathbf{\Delta \theta}} _{il} =  - \left ( 1- \frac{D_l}{D_i} \right ) \frac{4G_N}{c^2} \left\{ \frac{M^{\prime\prime}(b_{il}) \dot{|b_{il}|}^2}{b_{il}}\hat{\mathbf{ b}}_{il} \right.\\
 +  \frac{M^{\prime}(b_{il})}{b_{il}^2} \left[2 \dot{|b_{il}|} \mathbf{v}_{il}  + \ddot{|b_{il}|}\mathbf{ b}_{il} - 4 \dot{|b_{il}|}^2\hat{\mathbf{ b}}_{il} \right] \\
\left.  - \frac{2 M(b_{il})}{b_{il}^3}\left[2 \dot{|b_{il}|} \mathbf{v}_{il}  + \ddot{|b_{il}|}\mathbf{ b}_{il} + 3 \dot{|b_{il}|}^2\hat{\mathbf{ b}}_{il}\right] \right \}
\label{eq:acc}
\end{multline}
with $M^{\prime\prime}$ the second derivative of the enclosed mass with respect to the impact factor and 
\begin{equation} 
 \ddot{|b_{il}|} = \frac{\mathbf{v}_{il} \cdot \left(b_{il}\mathbf{v}_{il} -  \dot{|b_{il}|} \mathbf{ b}_{il} \right)}{b_{il}^2}.
\label{eq:bmagderder}
\end{equation}

In order to compute the deflection angle, velocity and acceleration (equations \ref{eq:deflang}, \ref{eq:vel} and \ref{eq:acc}) we  need to specify the distribution of matter in the lens, i.e., lens density profile $\rho(r)$. We  assume that the density profile of the halos is described by the Navarro–Frenk–White (NFW) profile \cite{1996ApJ...462..563N}
\begin{equation} 
\rho(r) = \frac{\rho_s}{\left(\frac{r}{r_s}\right)\left(1+\frac{r}{r_s}\right)^{2}}
\label{eq:rho_sb}
\end{equation}
where $r_s$ is the characteristic scale and $\rho_s$ the characteristic density of the profile.

\begin{figure*}
\centering
\includegraphics[width=2\columnwidth]{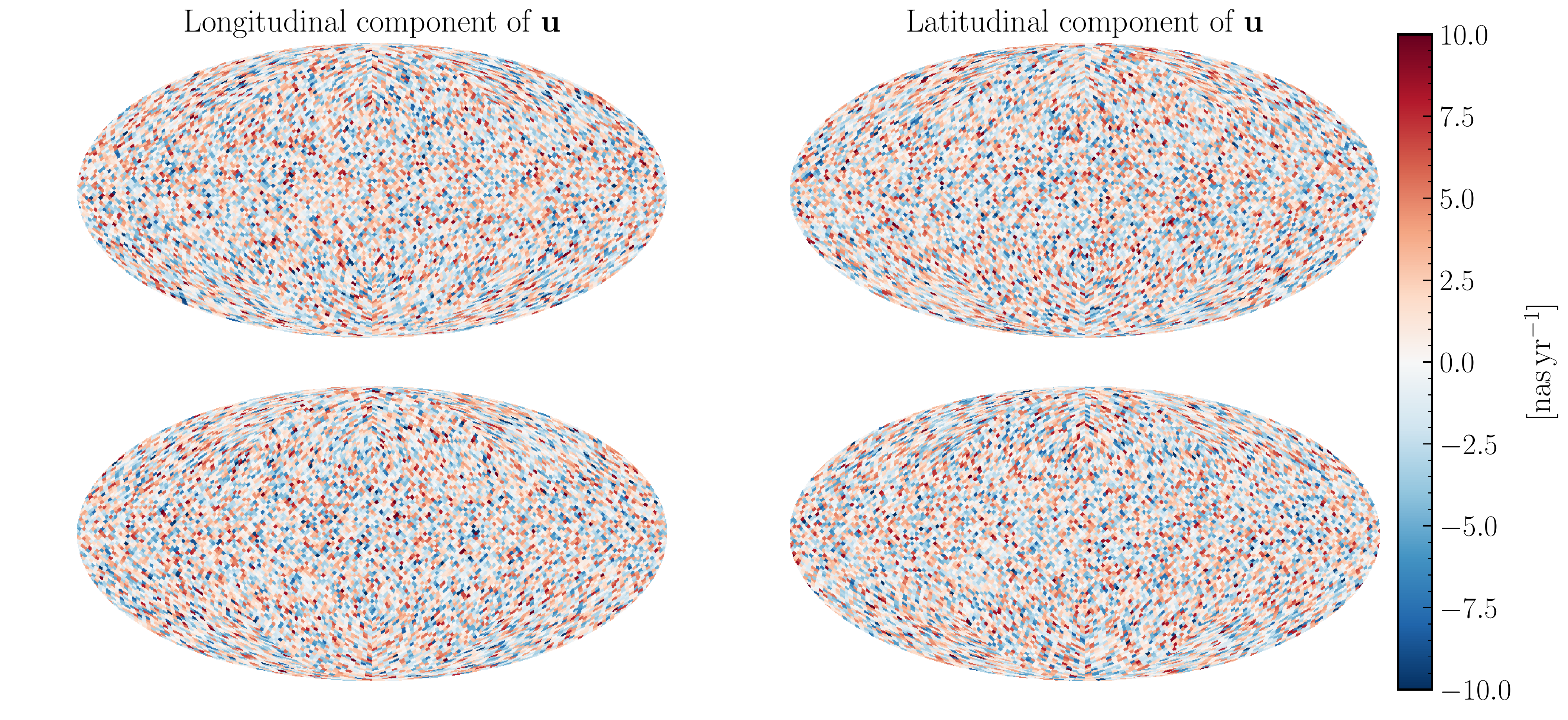}  
\caption{Top: The induced astrometric lensing velocity on a population of quasar sources by a population of subhalos distributed in the Milky Way superimposed with Gaussian noise based on the astrometric parameters $N_q = 10^9$ and $\sigma_{u} = 1~\rm{\mu as/yr}$ (Scenario C in Table 1). Bottom: Similar to top but with no signal (only noise).}
\label{fig:fig4}
\end{figure*}

In order to demonstrate the aforementioned effects, Figure~\ref{fig:fig2} shows an example of induced deflections, velocities and accelerations in arbitrary units (see also \cite{2018JCAP...07..041V}).  The lens is located at the origin moving towards the  $+\hat{\bf{y}}$ direction in all three plots. 
The left panel depicts the angular deflection.  The maximum angular deflection occurs for impact parameters approximately equal to the characteristic scale, as for $b_{il} < r_s$ the enclosed mass $M(<b_{il})$ decreases rapidly. On the other hand, if $b_{il} \gg r_s$ the enclosed mass approaches a constant and therefore the deflection decreases as $\sim 1/b_{il}$. The middle panel in Figure~\ref{fig:fig2} shows the corresponding induced velocity; the largest amplitude in the velocity vector occurs near the centre and in a direction opposite to the direction of motion of the lens with a dipole-like behavior in the far-field limit.   Finally, the right panel of Figure~\ref{fig:fig2} shows the acceleration pattern which exhibits a quadrupole-like behavior in the far-field.\footnote{Note that the amplitude of velocity and acceleration decreases faster than  the displacement field as each one of these picks an additional factor of $1/b_{il}$.} All these characteristics are in agreement to what has been demonstrated in of \cite{2018JCAP...07..041V}.

\subsection{Population of  lenses}
We now proceed to simulate the lensing effect on a population of sources by a population of lenses. Following \cite{2020arXiv200302264M}, we choose to simulate the lens population as NFW subhalos in the Milky Way that follow a mass function of the form 

\begin{equation} 
\frac{dN}{dM} = A_0M^{-1.9}
\label{eq:dNdM}
\end{equation}
\noindent
normalized so that there are 150 subhalos in the mass range $10^8 - 10^{10} M_\odot$ \cite{2016JCAP...09..047H} consistent with results from recent hydrodynamical simulations \cite{2015MNRAS.447.1353M,2016MNRAS.457.1931S}. 

We then define the spatial distribution of the subhalos, taking into account the tidal disruption due to the gradient of the Galactic potential towards the Galactic centre that depletes the fraction of mass bound in substructures in this region, by assuming an Einasto profile fitted to the results of the Aquarius simulation \cite{2012JMPh....3.1152R}
\begin{equation} 
\rho_g(r)=\exp\left\{-\frac{2}{\gamma_\varepsilon}\left[\left(\frac{r}{r_\varepsilon}\right)^{\gamma_\varepsilon}-1\right]\right\}.
\label{eq:einasto}
\end{equation}
\noindent
with $r_\varepsilon = 199\rm{kpc}$ and $\gamma_\varepsilon = 0.678$. We use this distribution to sample the radii at which the subhalos lay.

The next ingredient for defining a lens population is the dark matter velocity distribution. We assume that in the Galactic frame and far away from the Sun's gravitational potential the dark matter velocity distribution is described by the Standard Halo Model (SHM) as
\begin{equation} 
f_\infty(\mathbf{v}) = N \left(\frac{1}{\pi v_0^2}\right)^{3/2}e^{-\mathbf{v}^2/v_0^2}
\label{eq:SHM}
\end{equation}
\noindent
for $|\mathbf{v}|<v_{\mathrm{esc}}$ and $0$ otherwise (however see also \cite{2018PhRvD..97j3003V}). $N$ is a normalization factor to account for the truncation at $v_{\mathrm{esc}}$, $v_0=220 {\mathrm{km/s}}$ \cite{1986MNRAS.221.1023K} and the escape velocity is $v_{\mathrm{esc}}=550 {\mathrm{km/s}}$ \cite{2014A&A...562A..91P}. We transform this velocity distribution at the position of Earth (as we are interested in measurements at the Earth's frame) by applying a Galilean transformation from the Galactic frame to the Earth  frame 
\begin{equation} 
f_\oplus(\mathbf{v}) \approx f_\infty(\mathbf{v}+\mathbf{v}_\odot)
\label{eq:SHM_earth}
\end{equation}
\noindent
where $\mathbf{v}_\odot = \{11,232,7\}~\rm{km}/\rm{s}$ is the velocity of the Sun in  $\{U_\odot,V_\odot,W_\odot\}$ coordinates \cite{2010MNRAS.403.1829S}. We ignore the annual modulation due to the motion of the Earth around the Sun as it is a factor of ten smaller compared to the velocity components. 

We normalize the density profile of each subhalo in the following way. The NFW profile is defined by two parameters, $\rho_s$ and $r_s$, or alternatively, the mass of the halo, $M_{200}$ and its concentration $c_{200}$. The concentration is a measure of the compactness of the halos, defined as $c_{200} = R_{200}/r_s$, where $R_{200}$ is the radius that encloses $M_{200}$ whose density is 200 times the critical density of the universe $\rho_c=3H_0^2/8\pi G_N$. With this definition, the characteristic density $\rho_s$ and scale radius $r_s$ are obtained from 
\begin{eqnarray} 
M_{200} &=& \int_0^{r_{200}}dr^{\prime}4\pi r^{\prime 2} \rho(r^{\prime}) \nonumber \\
&=& 4 \pi \rho_s r_s^3 f(c_{200})
\label{eq:M_enc}
\end{eqnarray}
\noindent
where 
\begin{equation} 
f(x)=\ln(1+x) - \frac{x}{1+x} \nonumber
\label{eq:f}
\end{equation}
and  \cite{2014MNRAS.442.2271S} 
\begin{equation} 
c_{200}(M_{200}, z=0) = \sum_{i=0}^5 c_i \times \left[ \ln \left(\frac{M_{200}}{h^{-1}M_\odot}\right)\right]^i
\label{eq:c200}
\end{equation}
\noindent
with the factors $c_i$ having values $c_i=[37.5153, -1.5093, 1.636 \cdot 10^{-2} , 3.66 \cdot 10^{-4} ,
-2.89237 \cdot 10^{-5} , 5.32 \cdot 10^{-7} ]$.

It is clear that this choice of the description of subhalo profile parameters is unlikely as subhalos are evolved systems in the potential well of the Milky Way, while the prescription we use refers to field halos. Nevertheless, we make this choice not because it is realistic but because it can serve as an example to the proof of principle concept we explore here and also because it leads to a maximal signal. We plan to explore a thorough study of multiple other prescriptions of the mass distribution characteristics of substructure in future work.

\subsection{Population of sources}

We generate a population of sources with $D_i =\infty$ and $v_i=0$ at the center of each pixel of a HEALPIX map with ${\tt nside} = 128$, and we simulate  $\sim1200$ subhalos between $10^7$ and $10^{10} M_{\odot}$. This approach is similar to future observations of quasars with  the Square Kilometer Array (SKA), or other  instruments with similar capabilities. In the absence of instrumental noise, the resulting signal within 20 degrees from the center of each halo is shown in shown in \Fig{fig:fig3}. 

In  practice the sensitivity and precision of the instruments play a significant role. For this proof of principle example, we assume an SKA-like sensitivity\cite{ska}. Given  the current forecasts \cite{ska} after 10 years of operation SKA will be able to observe up to $N_q\sim 10^8$ quasars with a peculiar velocity uncertainty of $\sigma_{u}\approx 1~\mu as/{\mathrm{yr}}$ which we label scenario A. Additionally we consider four more sets of parameters (scenarios B, C, D and E) as shown in Table~\ref{tab:cases}. As we will see the weakness of the signal makes the significance of detection highly  dependent on these two parameters $N_q$ and $\sigma_{u}$.

\setlength{\tabcolsep}{22.5pt}
\renewcommand{\arraystretch}{1.3}

\begin{table}[h!]
\centering

\begin{tabular}{ |c|c|c|c| } 
\hline
Scenario & $N_q$ & $\sigma_{u} \ [\rm{\mu as/yr}]$\\
\hline
A & $10^8$ & 1 \\ 
B & $10^8$ & 0.1 \\ 
C & $10^9$ & 1 \\ 
D & $10^9$ & 0.1 \\ 
E & $3\times 10^9$ & 1 \\ 
\hline

\end{tabular}
\caption{Scenarios of astrometric parameters $N_q$ and  $\sigma_{u}$ used in this work.}
\label{tab:cases}
\end{table}

\section{The astrometric lensing signal}
\label{sec:DL}

\subsection{Characteristics of the signal in simulated\\ data sets}
In \Fig{fig:fig3} we show  the expected astrometric signal for the ideal case with no instrumental noise. We observe  qualitative characteristics similar to \Fig{fig:fig2}, such as the gradient of the magnitude of the displacement field as a function of distance is small compared to the gradient in the velocity and acceleration fields (due to the $1/b_{il}$ factors -- see footnote 1).  In addition, the colorbar depicts the magnitude of the expected effect. For the displacement field the median amplitude is $1~ \mathrm{mas}$, while for the velocity and acceleration fields the median values are $10^{-2} ~\mathrm{nas \  yr^{-1}}$, $5 \times 10^{-10}~\mathrm{nas \  yr^{-2}}$ respectively. 

For the purposes of this work, the velocity vector $\mathbf{u}$ is an adequate probe of substructure lensing as quasars are stationary in the galactic frame. However note that the  acceleration can also be used for some halo mass and concentration functions as was shown in \cite{2018JCAP...07..041V}.

In order to apply machine learning methods in our efforts to extract dark matter substructure from the astrometric data we first need to create a library of fake data sets which we will use in order to test the method outlined above. While \Fig{fig:fig3} shows the pure signal effect we must take into account instrumental noise. To create  simulated data sets we distribute the $N_q$ quasars on a HEALPIX grid with ${\tt{nside}} = 32$  corresponding to 12,288 equal-area pixels. We assume quasars follow a Poisson distribution among pixels with a mean value given by the total number of quasars divided by the total number of pixels. 

At each pixel we sample a longitudinal and latitudinal velocity component from a Normal distribution centered at the expected value in the presence of lensing and with a standard deviation  $\sigma_{u}$. With all quasars being assigned a measured velocity under the assumption that all quasars in a pixel will experience the same lensing effect, we average all quasars and assign a single velocity to each pixel. For the case of only instrumental noise, i.e., no signal, we follow the same procedure with the normal distribution always centered at 0. We repeat the process for all five scenarios of \Tab{tab:cases}.

\Fig{fig:fig4} shows an example of the simulated data after taking into account the sensitivity of the instrument for parameter set C. The upper two projections contain the lensing signal while the bottom two projections show the no-signal, only instrumental noise case. 

Next, we will use this method to generate big data sets for machine learning training and validation data sets. The hope is that by training a model to distinguish between a Gaussian noise from the non-Gaussian signal maps it will be possible to extract information about the underlying population of lenses in future surveys.

\begin{figure}
    \centering
    \includegraphics[width=\linewidth]{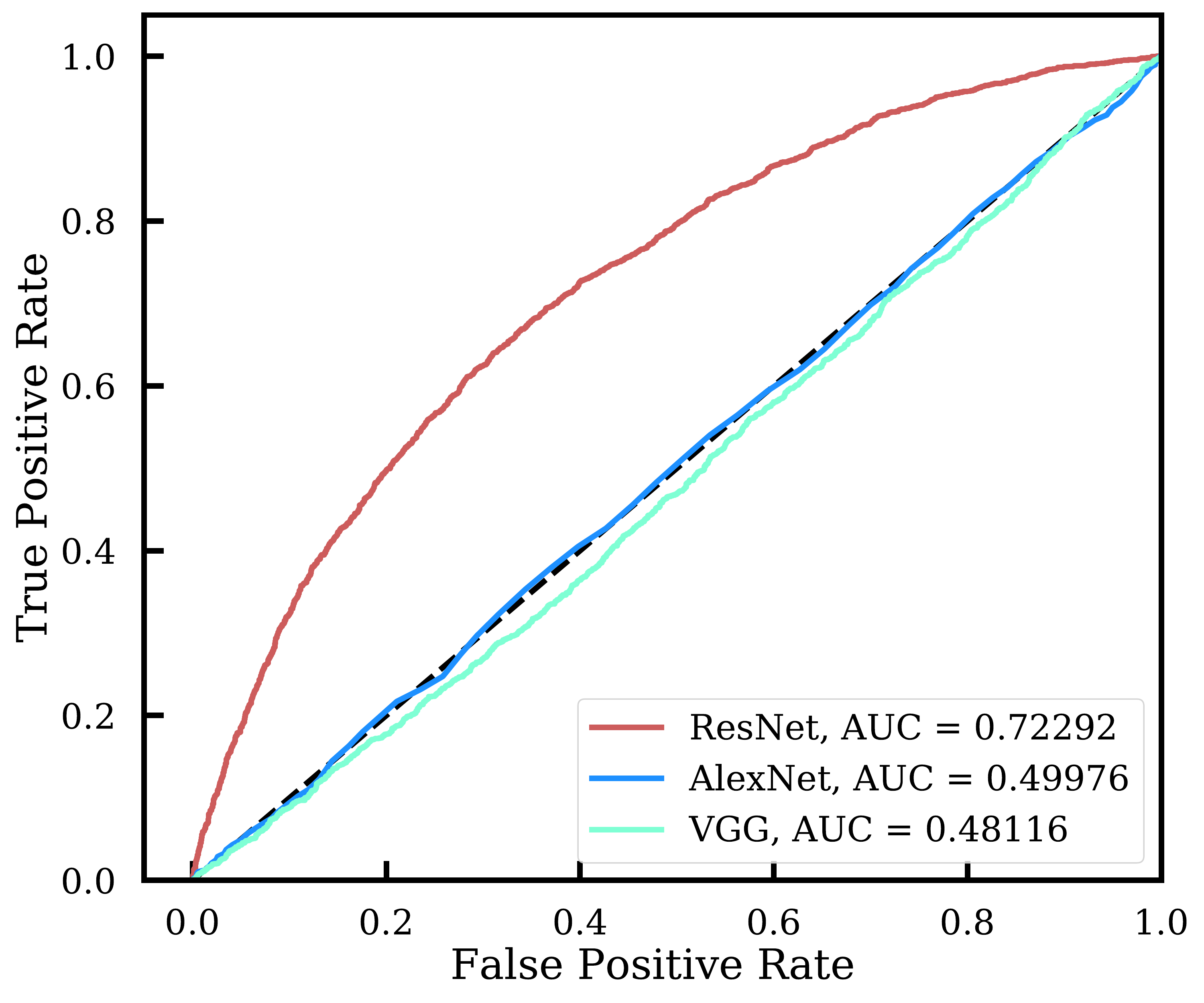}
    \caption{ROC curves of architecture performance between {\it ResNet-18}, {\it AlexNet}, and {\it VGG-11} for set C.  The black dashed line indicates equal True Positive and False Positive Rate corresponding to the case that a network is guessing in its classifications.}
    \label{fig:com-arc}
\end{figure}

\subsection{Machine learning implementation} 
The structure of the data leads naturally to an implementation of a CNN. One can appreciate this from \Fig{fig:fig4} where it is clear that correlations among pixels will likely serve a key role in helping to distinguish between the signal and no signal case. Given this intuition, we  trained state-of-the-art architecture \textit{ResNet-18} \cite{2015arXiv151203385H} as a classifier between these two classes. We selected \textit{ResNet-18} to train across all our sets after testing other algorithms ($VGG$ \cite{2014arXiv1409.1556S} and $AlexNet$ \cite{10.1145/3065386}), as it consistently achieved noticeably better performance - see  \Fig{fig:com-arc}.

\textit{ResNet-18} differs from a normal CNN in that it has residual blocks which are characterized by their use of skip connections, helping to alleviate the problem of vanishing gradients. One problem, however, is that our simulations are based on data in HEALPIX format. While there are architectures that are desinged to be trained on this data structure, notably graph based CNN DeepSphere \cite{Perraudin:2018rbt}, we instead opt to project our maps to Cartesian coordinates allowing us to train on a greater diversity of established architectures like \textit{ResNet}. 

For each scenario of Table 1, our training set consists of 25,000 training and 2,500 validation images per class (signal and no-signal cases). With a large number of simulated data, cross-validation is unnecessary as is the case for a lot of deep-learning applications. The cross-entropy loss was minimized over at most 50 epochs in batches of 32 with the Adam optimizer where the learning rate was initialized with a value of $1 \times 10^{-1}$ and decayed by a factor of 10 if the validation loss was not improved after 5 epochs. 

We implement {\it ResNet-18} with the \texttt{PyTorch} package and run on four NVIDIA Tesla V100 GPUs at the Brown Center for Computation and Visualization. As a metric for classifier performance we utilize the well-established Area Under the receiver operating characteristic (ROC) Curve (AUC).

\begin{figure}
    \centering
    \includegraphics[width=\linewidth]{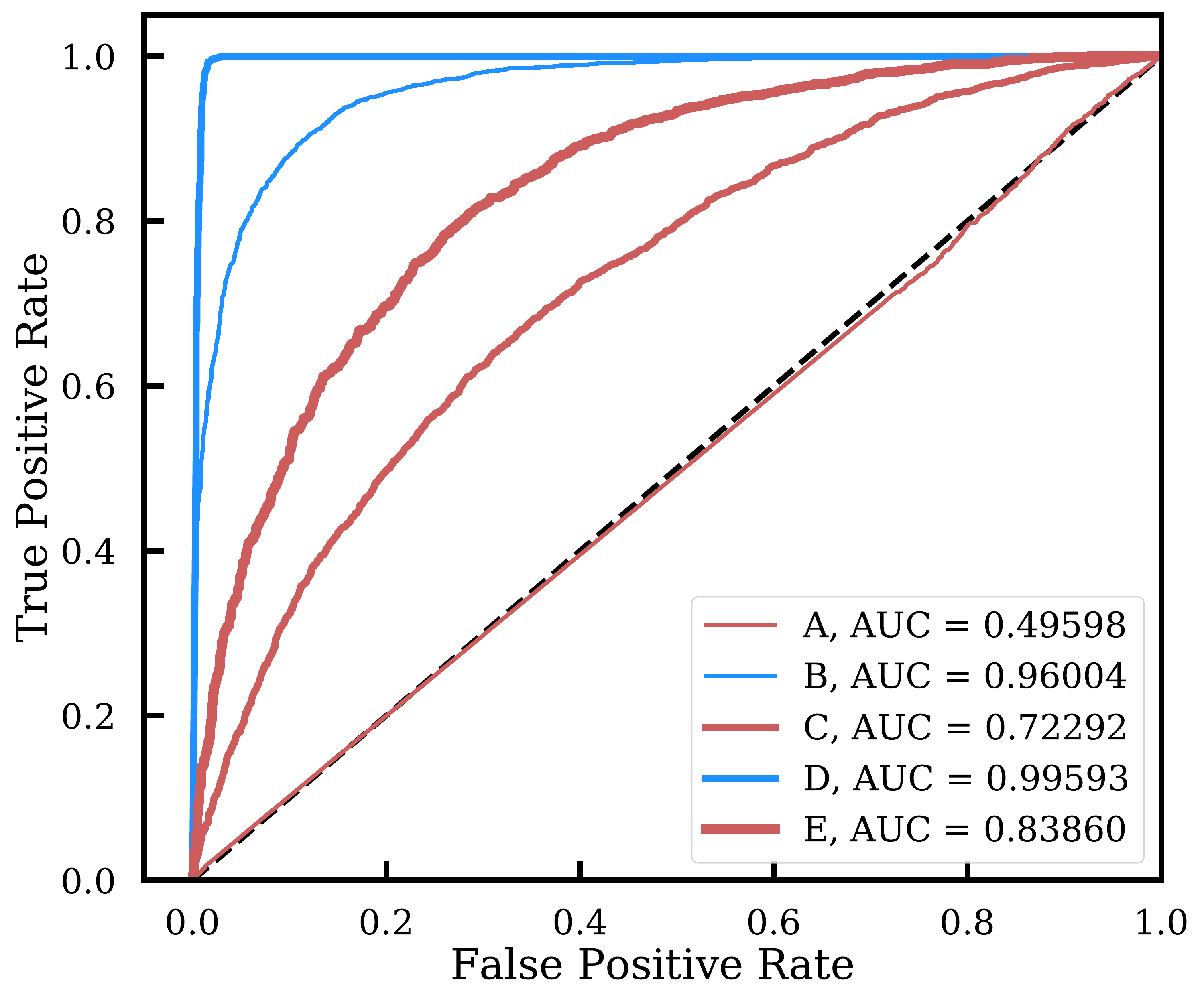}
    \caption{ROC curves and their corresponding AUC values for each set of values of $N_q$ and $\sigma_{u}$ as described in \Tab{tab:cases}. The red lines correspond to $\sigma_{u} = 1~\rm{\mu as/yr}$ while the blue lines are for $\sigma_{u} = 0.1~\rm{\mu as/yr}$ and the total number of quasars $N_q$ is designated by the line's width. The black dashed line indicates equal True Positive and False Positive Rate corresponding to the case that a network is guessing in its classifications.}
    \label{fig:ROC}
\end{figure}

\subsection{Integrated Gradients}
Neural networks are often referred to as {\it black boxes} - effectively a magician pulling a rabbit out of their hat. A true black box, however, would be of little interest. It is of critical importance if one is to use neural networks, say on real data, to garner some understanding of how the machine comes to making its decision.

One such method that has been developed recently is the method of attributing a networks prediction to its inputs. This is most frequently realized by analyzing gradients of the network output with respect to its input. A popular method is {\it integrated gradients} \cite{2017arXiv170301365S} which is realized as a path integral of gradients from a baseline to the desired input,
\begin{equation}
    IG_i(x,x') ::= (x_i - x_i') \int_{\alpha=0}^1 \frac{\partial F(x' + \alpha(x - x'))}{\partial x_i}d\alpha.
\end{equation}
Here $F$ corresponds to a trained model (perhaps a classifier between two types of dark matter) and $x$ an input (perhaps an astrometric lensing image velocity map). For a 2-dimensional image input, one can construct an assignment map where each pixel is assigned an attribution score. A positive (negative) attribution score contributes favorably (negatively) to the final network prediction whereas pixels with no attribution do not contribute. Inputs with the largest magnitude attribution score are the most influential in the networks final decision.

\begin{figure*}[t]
    \centering
    \includegraphics[width=0.9\linewidth]{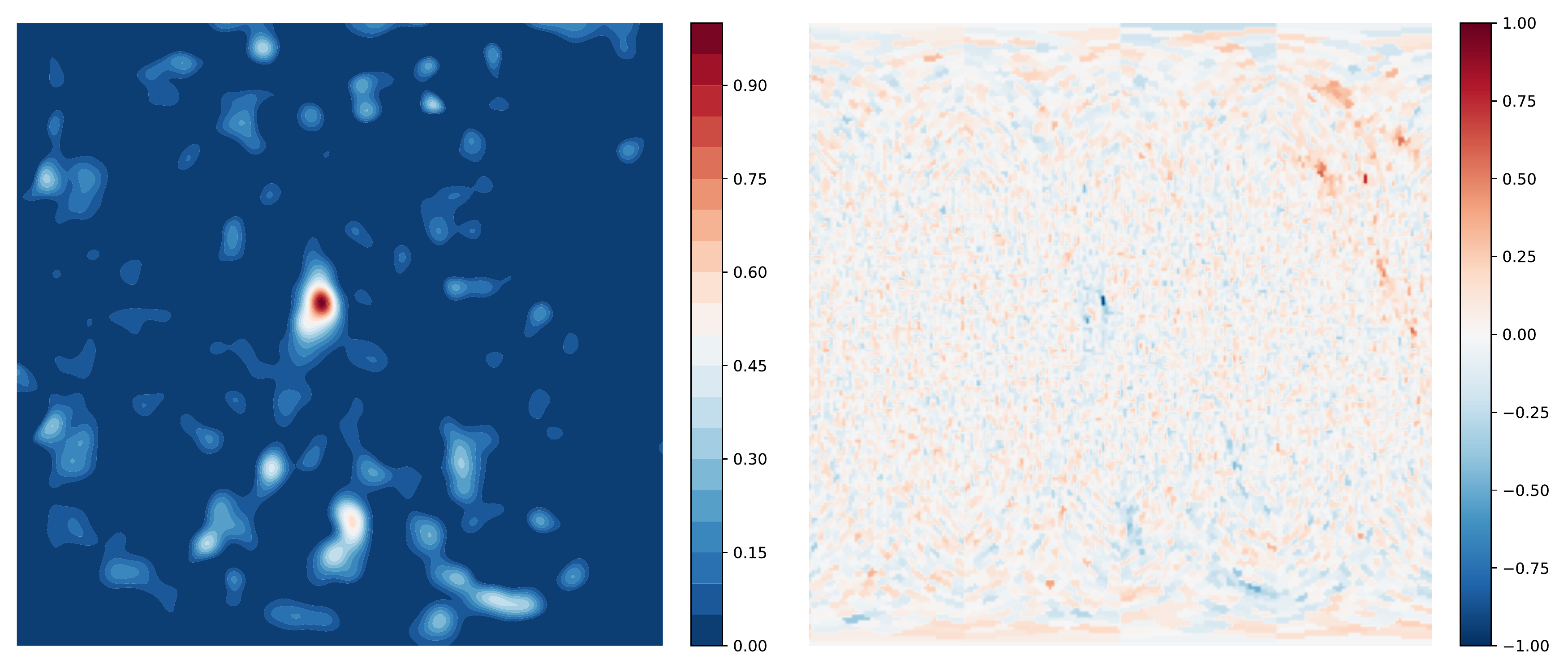}
    \caption{Example assignment map for {\it positive} attributions (left),  which has been convolved with a Guassian filter with $\sigma = 3$ pixels, calculated from a simulated lensing map with substructure from scenario D (right). The attribution score and astrometric lensing velocities have been normalized for convenience. The power of axiomatic attribution is now manifest in this example where it demonstrates that high, positive attribution scores can be useful in localization of substructure.}
    \label{fig:int-grad}
\end{figure*}

\section{Results}
\label{sec:results}

In \Fig{fig:ROC} we present the ROC curves and their corresponding AUC values for each set of parameters of \Tab{tab:cases}. This is the main result of this work.  The red lines correspond to $\sigma_{u} = 1~\rm{\mu as/yr}$ while the blue lines are for $\sigma_{u} = 0.1~\rm{\mu as/yr}$ and the total number of quasars $N_q$ is designated by the line's width. As expected larger $N_q$ and smaller $\sigma_{u}$ lead to higher AUC values and higher detection significance since the noise level of each pixel scales as $\sim \sigma_{u}/\sqrt{n_q}$ where $n_q$ is the number of quasars per pixel. Additionally we can see that a decrease in the uncertainty  $\sigma_{u}$ is more impactful than increasing the number of quasars $N_q$ and thus a more precise survey will be better suited to detect the signal than a more sensitive one.  

Our model trained on an SKA-like survey, realized here as parameter scenario A, is not sensitive to the signal induced by substructure after achieving an AUC of $\sim 0.5$, thus in this configuration the data is likely too noise-dominated even for a CNN. However, a more optimistic survey with ten times more quasars does manage to achieve limited power at discerning between signal and noise as evidenced by a marginal AUC score of 0.723 for scenario C, but this is still not very well equipped to detect dark matter substructure based on a 10 year operational baseline. 

In order to achieve a more robust performance, corresponding to AUC values $\gtrsim~0.8$, we need either the number of quasars to increase even more by a factor of at least 3 compared to scenario C, and/or $\sigma_{u}$ to decrease by $\sim$ a factor of 10, or some combination of the above. Encouragingly, it is very promising that the detection significance improves drastically with parameter scenario D where $N_q = 10^9$ and $\sigma_{u} = 0.1~\rm{\mu as/yr}$, allowing for almost perfect classification using the {\it ResNet-18} architecture. 

In other words, if SKA runs for longer than a decade, it is possible that the proposed technique can be successfully applied in the future to detect Milky Way substructure. For example, doubling its operational time reduces the uncertainty to $\sigma_{u} = 0.5~\rm{\mu as/yr}$, and with the currently planned $N_q \sim 10^8$ quasars it is possible to obtain an AUC of $\gtrsim 0.8$, and thus reliably probe the substructure content of the Milky Way. 

With several  classifiers in hand, it is possible to peer into the \textit{black box} using integrated gradients and extract information beyond our classification label. In Figure~\ref{fig:int-grad} we show the assignment map for {\it positive} attributions calculated from an example lensing map with substructure from scenario D. Since positive attributions correspond to pixels that voted for the substructure class, in principle they can be used to localize substructure in a lensing map.


Figure ~\ref{fig:int-grad} demonstrates the application of integrated gradients by showing that high attributions are correlated with substructure locations. Not surprisingly, it is prohibitively difficult to discern such features for data sets with lower AUC values as the usefulness of an attribution map is correlated with architecture performance.

We would like to point out that this method for substructure localization supplements other techniques that could be applied like image segmentation -- for example in the context of strong lensing see \cite{Ostdiek:2020cqz}  -- but should prove to be computationally more tractable since training is restricted to classification. As a final note, beyond the information contained in \textit{positive} attributions, features in the \textit{negative} attribution map should prove critical in understanding various noise and backgrounds in the data set in addition to improvements in the quality of data that may come by  longer observation times.

\section{Conclusion} \label{sec:conclusion}
We investigate the possibility of using supervised deep learning techniques to detect the astrometric lensing effect on a population of quasars from dark matter subhalos in the Milky Way as a proof of principle using simulated data. In particular, we apply state-of-the-art convolutional neural network {\it ResNet-18} to our data set and we show how a CNN could be used to detect the astrometric lensing signal from future astrometric surveys provided an architecture trained on simulations were to be applied to a real data set.

We observe that for the current best velocity sensitivity estimates for a 10 year run of an SKA-like survey $\sigma_{u} \sim 1~\rm{\mu as/yr}$, a total number of $N_q \sim 3 \times 10^9$ quasars would be required for the classifier to detect the signal with an AUC value $>0.8$. Much better performance is achieved for $N_q \sim 10^8$ but much smaller velocity uncertainty $\sigma_{u} \sim 0.1~\rm{ \mu as/yr}$ with AUC value of $\sim 0.96$. We see that a smaller uncertainty has much higher impact on the ability of the classifier to detect the dark matter substructure and it can be achieved by expanding the duration of the survey to more than a decade. Our results are in qualitative agreement with the results of \cite{2020arXiv200302264M} though we cannot quantitatively compare since the AUC values used here don't directly map to the discovery significance shown in Fig. 6 of that work.

We have further shown a method for localizing subhalos using integrated gradients, a method of axiomatic attribution. Concretely, we find that the map for \textit{positive} attributions, pixels which voted for the substructure class, are found to be consistent with the location of known strong lenses in our simulations. While we don't investigate in detail here, the \textit{negative} attributions are likely to encode crucial information about various noise and backgrounds present in the data set.

While it may be possible that there is an architecture that could be better suited to tackle the complexity of the data set at the level of expectations for SKA, it should be pointed out that the results presented here are based upon an assumed dark matter model defined by the mass and distribution of its substructure. It may be the case that dark matter is something other than the CDM picture presented here and, thus, prospects for detection may differ. Similar to the spirit of \cite{alexander2020lensing} where anomaly detection, a form of \textit{unsupervised} machine learning, was used to detect the presence of dark matter substructure in strong lensing images, it seems natural to extend this work to the unsupervised scenario where an autoencoder is trained on sets of images corresponding to SKA measurements with no substructure and qualifying the constraining power of identifying any DM substructure in SKA data. In this way one is taking a \textit{theory agnostic} approach to the detection of dark matter – i.e. our classifier performance will not depend on an arbitrary choice of model and its various parameters. Additionally the inclusion of acceleration data on top of the velocity maps can further strengthen the detection of the signal. We leave this for future work.

\acknowledgements
We acknowledge useful conversations with Stephon Alexander, Jatan Buch, and John Leung. This research  was conducted using computational resources and services at the Center for Computation and Visualization, at Brown University. SMK is partially supported by NSF PHY-2014052. We gratefully acknowledge the support of Brown University.

\bibliography{manuscript}

\end{document}